\documentclass[%
 preprint,
 superscriptaddress,
 amsmath,amssymb,
 aps,
prd,
 floatfix,
]{revtex4-1}
\usepackage{graphicx} 
\usepackage{amsfonts}
\usepackage{float}
\usepackage{physics}
\usepackage{xcolor}
\usepackage{siunitx}
\usepackage{comment}
\usepackage{hyperref}

\begin{document}
\title{Search for high-frequency gravitational waves via re-analysis of cavity axion data}

\author{Younggeun Kim}
\email{kimyoung@uni-mainz.de}
\affiliation{Johannes Gutenberg-Universit{\"a}t Mainz, 55122 Mainz, Germany}
\affiliation{Helmholtz Institute Mainz, 55128 Mainz, Germany}
\affiliation{GSI Helmholtzzentrum für Schwerionenforschung GmbH, 64291 Darmstadt, Germany}
\author{Jordan Gué}
\affiliation{Institut de F\'{i}sica d’Altes Energies (IFAE), The Barcelona Institute of Science and Technology, Campus UAB, 08193 Bellaterra (Barcelona), Spain}
\author{Changhao Xu}
\affiliation{Johannes Gutenberg-Universit{\"a}t Mainz, 55122 Mainz, Germany}
\affiliation{Helmholtz Institute Mainz, 55128 Mainz, Germany}
\affiliation{GSI Helmholtzzentrum für Schwerionenforschung GmbH, 64291 Darmstadt, Germany}
\author{Diego Blas}
\affiliation{Institut de F\'{i}sica d’Altes Energies (IFAE), The Barcelona Institute of Science and Technology, Campus UAB, 08193 Bellaterra (Barcelona), Spain}
\affiliation{{Instituci\'{o} Catalana de Recerca i Estudis Avan\c{c}ats (ICREA),
Passeig Llu\'{i}s Companys 23, 08010 Barcelona, Spain}}
\author{Dmitry Budker}
\affiliation{Johannes Gutenberg-Universit{\"a}t Mainz, 55122 Mainz, Germany}
\affiliation{Helmholtz Institute Mainz, 55099 Mainz, Germany}
\affiliation{GSI Helmholtzzentrum für Schwerionenforschung GmbH, 64291 Darmstadt, Germany}
\affiliation{Department of Physics, University of California, Berkeley, CA 94720-7300, United States of America}
\author{Sungjae Bae}
\affiliation{Dark Matter Axion Group, Institute for Basic Science, Daejeon 34126 Republic of Korea}
\affiliation{Center for Axion and Precision Physics Research, Institute for Basic Science, Daejeon 34051, Republic of Korea}
\author{Claudio Gatti}
\affiliation{NFN Laboratori Nazionali di Frascati, via Enrico Fermi 54, 00044 Frascati (Roma), Italy}
\author{Junu Jeong}
\affiliation{Oskar Klein Centre, Department of Physics, Stockholm University, AlbaNova, SE-10691 Stockholm, Sweden}
\author{Jihn E. Kim}
\affiliation{Department of Physics, Seoul National University, Seoul 08826, Republic of Korea}
\author{Kiwoong Lee}
\affiliation{Center for Axion and Precision Physics Research, Institute for Basic Science, Daejeon 34051, Republic of Korea}

\author{Arjan F. van Loo}
\affiliation{RIKEN Center for Quantum Computing, Wako, Saitama 351-0198, Japan}
\affiliation{Department of Applied Physics, Graduate School of Engineering, The University of Tokyo, Bunkyo-ku, Tokyo 113-8656, Japan}
\thanks{Current address: Alice $\&$ Bob, 49 Boulevard Du Général Martial Valin, Paris, 75015, France}
\author{Yasunobu Nakamura}
\affiliation{RIKEN Center for Quantum Computing, Wako, Saitama 351-0198, Japan}
\affiliation{Department of Applied Physics, Graduate School of Engineering, The University of Tokyo, Bunkyo-ku, Tokyo 113-8656, Japan}
\author{Seonjeong Oh}
\affiliation{Dark Matter Axion Group, Institute for Basic Science, Daejeon 34126 Republic of Korea}
\affiliation{Center for Axion and Precision Physics Research, Institute for Basic Science, Daejeon 34051, Republic of Korea}
\author{Wolfram Ratzinger}
\affiliation{Department of Particle Physics and Astrophysics, Weizmann Institute of Science,
Herzl Street 234, Rehovot, 7610001, Israel}
\author{Taehyeon Seong}
\affiliation{Dark Matter Axion Group, Institute for Basic Science, Daejeon 34126 Republic of Korea}
\affiliation{Center for Axion and Precision Physics Research, Institute for Basic Science, Daejeon 34051, Republic of Korea}
\author{Yannis K. Semertzidis}
\affiliation{Center for Axion and Precision Physics Research, Institute for Basic Science, Daejeon 34051, Republic of Korea}
\affiliation{Department of Physics, Korea Advanced Institute of Science and Technology, Daejeon 34141, Republic of Korea}

\author{Kristof Schmieden}
\affiliation{Universit{\"a}t Bonn, 53113, Bonn, Germany}
\author{Mattias Schott}
\affiliation{Universit{\"a}t Bonn, 53113, Bonn, Germany}
\author{Sergey Uchaikin}
\affiliation{Dark Matter Axion Group, Institute for Basic Science, Daejeon 34126 Republic of Korea}
\affiliation{Center for Axion and Precision Physics Research, Institute for Basic Science, Daejeon 34051, Republic of Korea}
\author{SungWoo Youn}
\affiliation{Dark Matter Axion Group, Institute for Basic Science, Daejeon 34126 Republic of Korea}
\affiliation{Center for Axion and Precision Physics Research, Institute for Basic Science, Daejeon 34051, Republic of Korea}
\begin{abstract}
    Monochromatic high-frequency gravitational waves (HFGW) provide a distinctive probe of new physics scenarios, most notably axion clouds around rotating black holes formed via superradiance. We reanalyzed data from the CAPP–12T MC (multi-cell) axion haloscope experiment [Phys. Rev. Lett. 133,051802 (2024)]. The study covers a continuous 2\,MHz frequency span centered at 5.311\,GHz. No rescan candidates were found, and we set 90\% confidence-level exclusion limits on the gravitational-wave strain, reaching $h_0 \approx 3.9 \times 10^{-21}$ in the most sensitive regions of the sky. Interpreted in the context of black-hole superradiance from axion clouds, the results exclude black holes with mass $M_{\mathrm{BH}} \simeq 1.22 \times 10^{-6}\,M_\odot$ within distances of $\mathcal{O}(10^{-2})$\,AU from Earth, under benchmark assumptions. This work demonstrates the potential of electromagnetic resonant cavities as novel detectors of monochromatic HFGW and motivates future searches for both long-lived and transient signals.
\end{abstract}

\maketitle

\section{Introduction}

Gravitational waves (GW) have been firmly established as a new observational probe of the Universe. Ground-based interferometers such as LIGO, Virgo, and KAGRA have directly detected compact binary mergers in the audio band (tens of Hz to several kHz)~\cite{LIGO2016, Virgo2015, KAGRA2020}, while pulsar timing arrays (PTA) have recently reported evidence for a nanohertz GW background~\cite{NANOGrav2023, EPTA2023}. At the lowest frequencies, cosmic microwave background (CMB) polarization places indirect limits on primordial GWs~\cite{Planck2018, BICEP2021}. Together, these observations demonstrate that the GW spectrum spans a broad frequency range, with each band probing distinct aspects of astrophysics and cosmology.
The frequency reach of current detectors, however, remains limited. Interferometers are confined to the Hz-kHz range due to seismic and quantum noise, PTA probe only the nHz regime, and the CMB constrains frequencies far below a Hz. Crucially, no known stellar or compact-object mechanism is expected to produce significant gravitational-wave emission in the MHz--GHz range. 
As summarized in recent literature and reviews~\cite{Caprini2018,Aggarwal:2025noe}, standard astrophysical sources are limited to $\lesssim\,$kHz frequencies, 
and the MHz--GHz window is instead most naturally motivated by cosmological and beyond-standard-model scenarios. Nevertheless, this high-frequency GW (HFGW) window is theoretically well motivated~\cite{Aggarwal:2025noe}: several cosmological scenarios predict radiation at such frequencies, including preheating after inflation~\cite{Khlebnikov1997}, first-order phase transitions~\cite{PhysRevLett.112.041301,PhysRevD.96.103520}, and cosmic string networks~\cite{Siemens2009}. In addition, signals from primordial black holes (chirping, or rather monochromatic, depending on the mass range) are possible \cite{Franciolini:2022htd} and monochromatic GW signals are predicted from axion clouds around rotating black holes through the process of superradiance~\cite{Arvanitaki2010, Arvanitaki2015, Brito2020}, providing a distinctive and coherent target for experiments. 

Experimental efforts to access this unexplored band have begun only recently. 
The MAGO project, based on superconducting coupled microwave resonators, enables HFGW searches over the 10\,kHz–1\,GHz range in both resonant and broadband schemes~\cite{Ballantini:2003aa,ballantini2005,PhysRevD.108.084058}. Meanwhile, ABRACADABRA-type experiments explore kHz–MHz magnetic pickup signatures~\cite{Valerie2022_PRL,Pappas:2025zld}.
Bulk acoustic wave (BAW) resonators~\cite{Goryachev2014,Goryachev2021,Campbell:2023aa,campbell2025_mage} and interferometric experiments such as the Fermilab Holometer~\cite{Chou2017} have also probed the MHz range. These pioneering attempts provide initial sensitivity, but the GHz domain remains largely unexplored~\cite{Aggarwal:2025noe}.
A particularly promising approach is to employ electromagnetic resonators, exploiting the inverse Gertsenshtein effect~\cite{Gertsenshtein1962}: a GW interacting with a static magnetic field induces effective currents that couple to resonant electromagnetic modes of a cavity~\cite{Cruise2000, Berlin2021,Ahn:PRD2024,Ratzinger:2024aa}. Resonant enhancement, combined with ultralow-noise readout, makes RF and microwave cavities natural HFGW detectors. 

Microwave cavities in axion haloscope experiments represent the state-of-the-art realization of this idea. These searches employ cryogenic high-$Q$ cavities in strong magnetic fields, read out with quantum-limited amplifiers, to probe the extremely weak couplings of dark-matter axions~\cite{Sikivie1983, ADMX2023, CAPP2024_12TMC,CAPP2024_12TBPRX,CAPP2025_12T_TM020,ORGANPRL2024,QUAX2024,HAYSTAC2025_PRL}. Beyond axions, the same hardware is ideally suited for narrow-band HFGW detection. Crucially, haloscopes have already accumulated extensive high-quality spectral datasets, which can be repurposed for GW analyses. The CAPP–12T MC haloscope~\cite{CAPP2024_12TMC} is among the most sensitive axion cavity experiments to date, operating near 5 GHz and achieving sensitivity below the benchmark KSVZ (Kim–Shifman–Vainshtein–Zakharov) QCD axion model~\cite{Kim:PRL:1979, Shiftman:NPB:1980}. Its multi-cell copper cavity, operated in a 12\,T superconducting solenoid and coupled to a Josephson Parametric Amplifier (JPA), achieved system noise near the quantum limit. Long integrations were performed across MHz-wide ranges, yielding data directly applicable to searches for long-lived monochromatic GWs.

In this work, we present a re-analysis of a subset of the CAPP-12T MC dataset, covering a 2\,MHz span centered at 5.311\,GHz. By exploiting the high quality factor, ultralow-noise performance, and extended integration of the experiment, we set new exclusion limits on monochromatic HFGW in the gigahertz band. Interpreted in terms of black-hole superradiance, these results exclude the existence of nearby light black holes surrounded by axion clouds within $\mathcal{O}(10^{-2})$\,AU, and demonstrate the broader scientific reach of haloscope experiments beyond their original dark-matter goal.

\section{Sources of monochromatic GW}\label{sec:GWources}

The sources of HFGW ($\gtrsim$ MHz) are generally expected to be either stochastic backgrounds or burst-like signals, such as those from preheating, phase transitions, or cosmic strings, producing broad spectral features. In contrast, \textit{monochromatic} sources are more exotic in this frequency range since sustaining a single frequency emission requires a highly coherent process. The leading candidate is axion-cloud annihilation generated via black-hole superradiance~\cite{Aggarwal:2025noe, Arvanitaki2010, Arvanitaki2015, Brito2020}. Primordial black-hole (PBH) mergers may also generate these monochromatic signals, but for masses far from the reach of current detectors \cite{Franciolini:2022htd}. For simplicity, we focus on the superradiant case, though our sensitivity reach can be rescaled for the latter. 

Superradiance, in the context of gravitational waves, refers to an enhanced radiation process associated with bosonic fields around rotating black holes. The event horizon of a spinning black hole is a particularly suitable system for this phenomenon~\cite{Arvanitaki2015}. 
The condition for superradiance is satisfied when $m_{a}$, the mass of the bosonic field (e.g. an axion), is matched to the oscillation frequency $\omega_{R}$ of the quasi-bound states of the bosonic particle in the gravitational potential~\cite{Aggarwal:2025noe}. 
Among the several possible emission channels, the dominant monochromatic GW process is axion annihilation ($a+a \rightarrow h$) where $a$ is the axion and $h$ is the graviton \cite{Arvanitaki:2010sy,Arvanitaki:2014wva,Zhu:2020tht}, leading to a GW frequency corresponding to twice the axion mass. For our data taken near 5.3\,GHz, we have
\begin{equation}
f_{\mathrm{GW}} = \frac{\omega_{G}}{\pi} = 5.3\,\mathrm{GHz}\left(\frac{m_{a}}{11\,\mu\rm{eV}}\right). 
\label{eq:ma_superradiance}
\end{equation}
The corresponding black hole mass $M_{\rm{BH}}$ can be evaluated from the typical value of $\alpha = GM_{\rm{BH}}m_{a}/\hbar c$. For the non-relativistic condition $\alpha<1$, if we choose $\alpha =0.1$~\cite{Aggarwal:2025noe}, the corresponding black hole mass is given as:
\begin{equation}
M_{\mathrm{BH}} = 1.2\times10^{-6}\,M_{\odot} \left(\frac{\alpha}{0.1}\right)\left(\frac{11\,\mathrm{\mu eV}}{m_{a}}\right).
\end{equation}
Then, the expected GW strain for axion annihliations is (for the case of a gravitational atom with an angular momentum of  $l =1$)~\cite{Aggarwal:2025noe,Arvanitaki:2014wva,Zhu:2020tht},
\begin{equation}
    h_{0} \approx 1.0 \times 10^{-22}\left(\frac{\alpha}{0.1}\right)^7\left(\frac{1\,\mathrm{AU}}{D}\right)\left(\frac{M_{\mathrm{BH}}}{ 10^{-6}M_{\odot}}\right)\,,
\end{equation}
where $D$ is the distance to the source in astronomical units (AU). The corresponding duration of this emission phase is~\cite{Aggarwal:2025noe},
\begin{equation}
    \tau_s \approx 4.7 \, {\rm days}  \left(\frac{\alpha}{0.1}\right)^{-15}\left(\frac{M_{\mathrm{BH}}}{ 10^{-6}M_{\odot}}\right).
\end{equation}
Hence, only black holes that have recently achieved spin are candidates for current searches. This is not a natural scenario for PBH.  Other processes may yield more long-lasting signals (for example, in the presence of self-interactions \cite{Yoshino:2015nsa,Arvanitaki:2014wva,Zhu:2020tht,Brito2020}).  Since the target of this work is determination of the sensitivity to $h_0$ for a given monochromatic HFGW lasting at least the data taking period, we do not further delve into the origin of the signal and let the reader translate our bounds to their favorite model.

\section{CAPP-12T MC Experimental Setup}

We reanalyze part of the dataset collected in the CAPP-12T MC axion haloscope experiment~\cite{CAPP2024_12TMC}. 
The experiment employs a multi-cell microwave cavity with an internal volume of $1.38\,\mathrm{L}$ (inner diameter: 78\,mm, height : 300\,mm), constructed from oxygen-free high-conductivity copper, placed inside a superconducting solenoid providing a $12\,\mathrm{T}$ magnetic field at the center of the magnet. 
The cavity was cooled down to below $40\,\mathrm{mK}$ with a dilution refrigerator, yielding a physical temperature well below the quantum limit of half a photon at the corresponding cavity frequency (130\,mK at 5.3\,GHz)~\cite{PhysRev.83.34}. During this run, the loaded quality factor was typically $Q_{l} \simeq 1.3\times10^{4}$.
\begin{figure}[h!]
    \centering
    \includegraphics[width=0.9\linewidth]{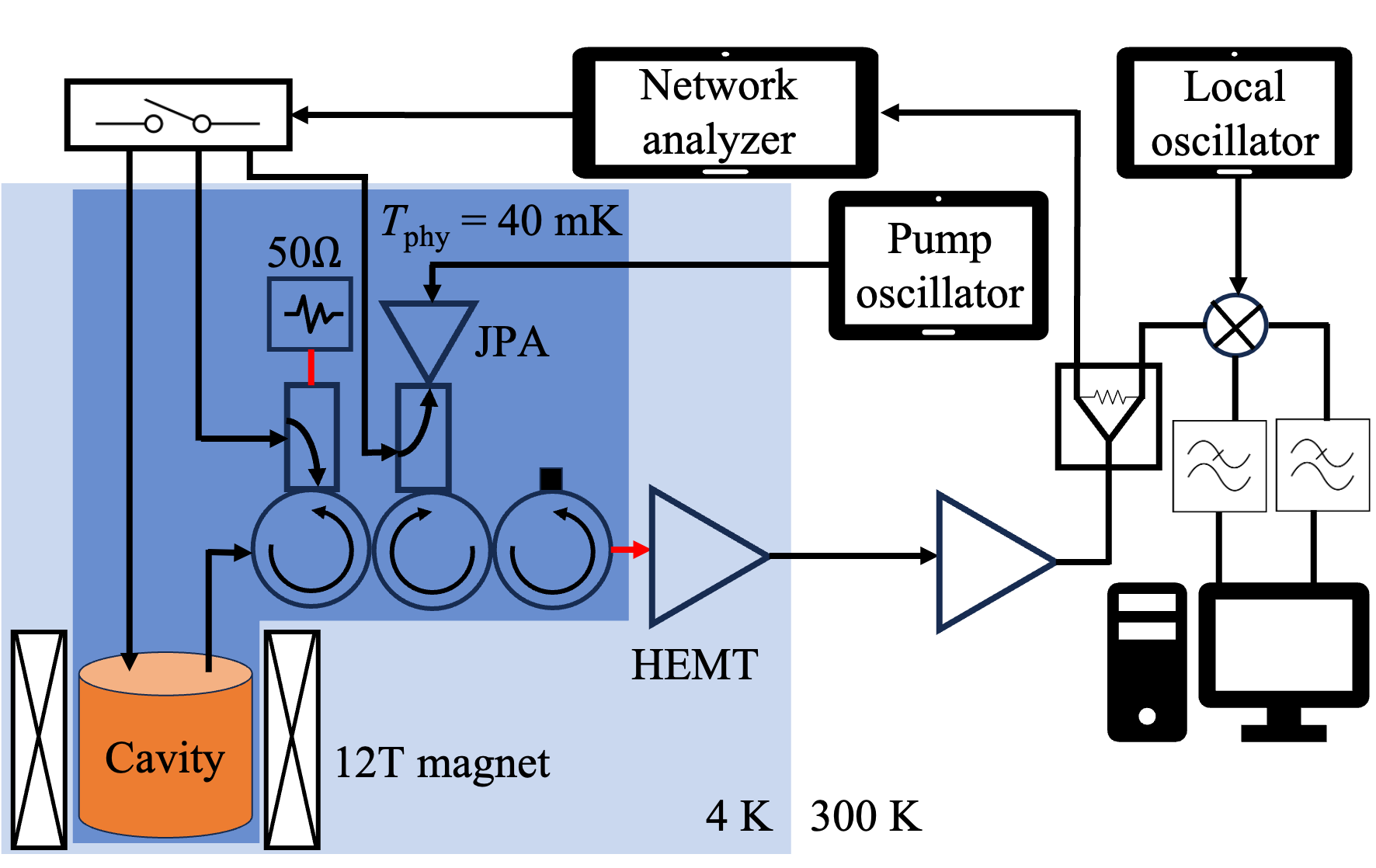}
    \caption{Schematic of the CAPP-12T MC haloscope setup~\cite{CAPP2024_12TMC}. Red solid lines (between the 50
    \,$\Omega$ noise source and the directional coupler, and between the isolator at mixing plate of physcial temperature 40\,mK and the HEMT amplifiers) indicate superconducting RF lines that prevent thermal links.}
    \label{fig:CAPP12T_scheme}
\end{figure}
A flux-driven Josephson Parametric Amplifier (JPA) was used as the first-stage amplifier, which is magnetically shielded with multiple layers of aluminum, cryoperm, and NbTi. 
The JPA is followed by two cryogenic High Electron Mobility Transistor (HEMT) amplifiers, with circulators and isolators placed in between to mitigate impedance mismatch between the amplifier stages. 
The system noise temperature, measured using the noise visibility ratio~\cite{NoiseFigure_1944,Sliwa_noise_PhysRevX}, was $T_{\mathrm{sys}} = 380\,\mathrm{mK}$ ($\simeq 1.5\,h\nu$ around $5\,\mathrm{GHz}$) on average, consistent with an independent measurement of the system noise temperature via the Y-factor~\cite{Pozar:882338}, within $2\%$.

The signal was mixed down to an intermediate frequency (IF) of $3\,\mathrm{MHz}$ using an IQ mixer, with the in-phase and quadrature channels processed independently before being digitized for 10\,s at a $20\,\mathrm{MHz}$ sampling rate and recombined in software. 
The resulting time-domain data were Fourier transformed to obtain spectra with $100\,\mathrm{Hz}$ resolution over a $1\,\mathrm{MHz}$ span around the IF, which were stored for subsequent analysis.

Data acquisition (DAQ) begins with the characterization of the cavity parameters using a network analyzer (NA).
The resonance frequency ($\nu_{c}$) and loaded quality factor ($Q_{l}$) are obtained by Lorentzian fits to the transmission trace, while the antenna coupling, $g$, is determined from Smith-chart fitting of the reflection trace~\cite{IEEE:Q_coupling:1984}.
The resonance of the Josephson parametric amplifier (JPA) is then tuned 200\,kHz below the cavity resonance to operate in the non-degenerate mode~\cite{Zhong_2013,Yamamoto2016} at the cavity resonance.
For each tuning step, the JPA operating point was automatically optimized \textit{in situ} using the Nelder–Mead algorithm~\cite{Nelder_Mead_1965,YKim2024_electronics} to ensure minimal noise temperature and stable gain.
At each tuning step, spectra were accumulated for 7-8 hours, and the cavity resonance was shifted in $150\,\mathrm{kHz}$ steps—smaller than half of the loaded bandwidth—to guarantee overlap between adjacent settings. 
During acquisition, the DAQ automatically paused every six minutes to check the JPA operating point and cavity response; if the gain varied by more than $1\,\mathrm{dB}$, the NM routine was re-run before continuing. 
Further details of the cavity geometry and design can be found in Ref.~\cite{CAPP2024_12TMC}.
The full data-taking run spanned 82 days, from July to November 2023. 
In this work, we focus on the subset of data collected between September 25 and 30, covering a continuous $2\,\mathrm{MHz}$ frequency band centered at $5.311\,\mathrm{GHz}$, where the system noise temperature reached its minimum of $360\,\mathrm{mK}$.

\section{Conversion power of gravitational wave in resonant cavity and gravitational coupling to the mode}

\begin{figure}[htbp]
\centering
    \includegraphics[width=0.7\linewidth]{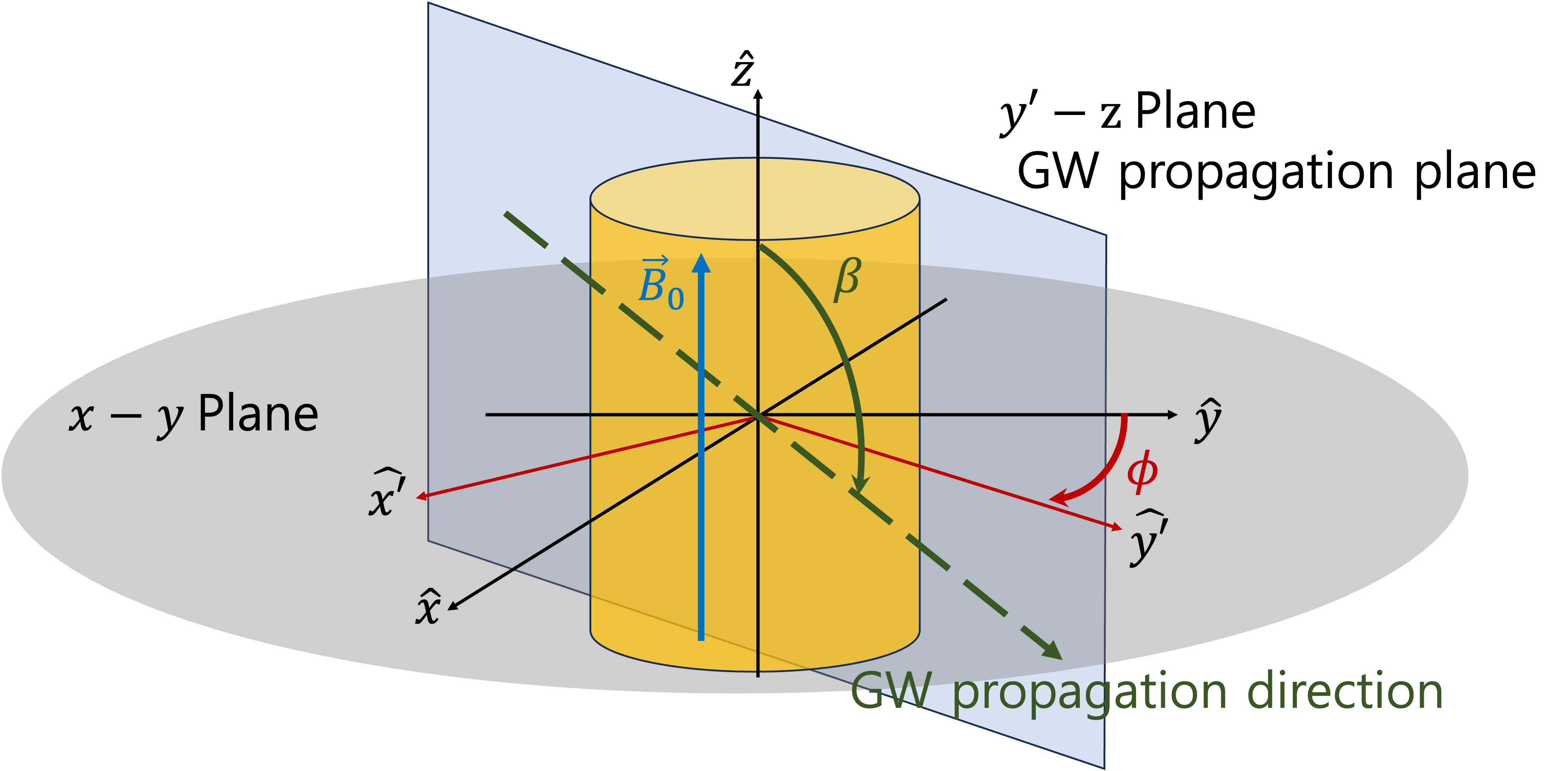}
    \caption{Schematic of the GW propagation plane with respect to the cylindrical cavity. The external magnetic field is applied along $\hat{z}$ axis. GW propagates in the $y’–z$ plane, making an angle $\beta$ with the cavity axis $\hat{z}$. The azimuthal angle $\phi$ defines the orientation of the GW propagation direction with respect to the $x–y$ plane.}
    \label{fig:GW_cav_angle}
\end{figure}

\begin{figure}[h!]
\centering
\includegraphics[width=0.9\linewidth]{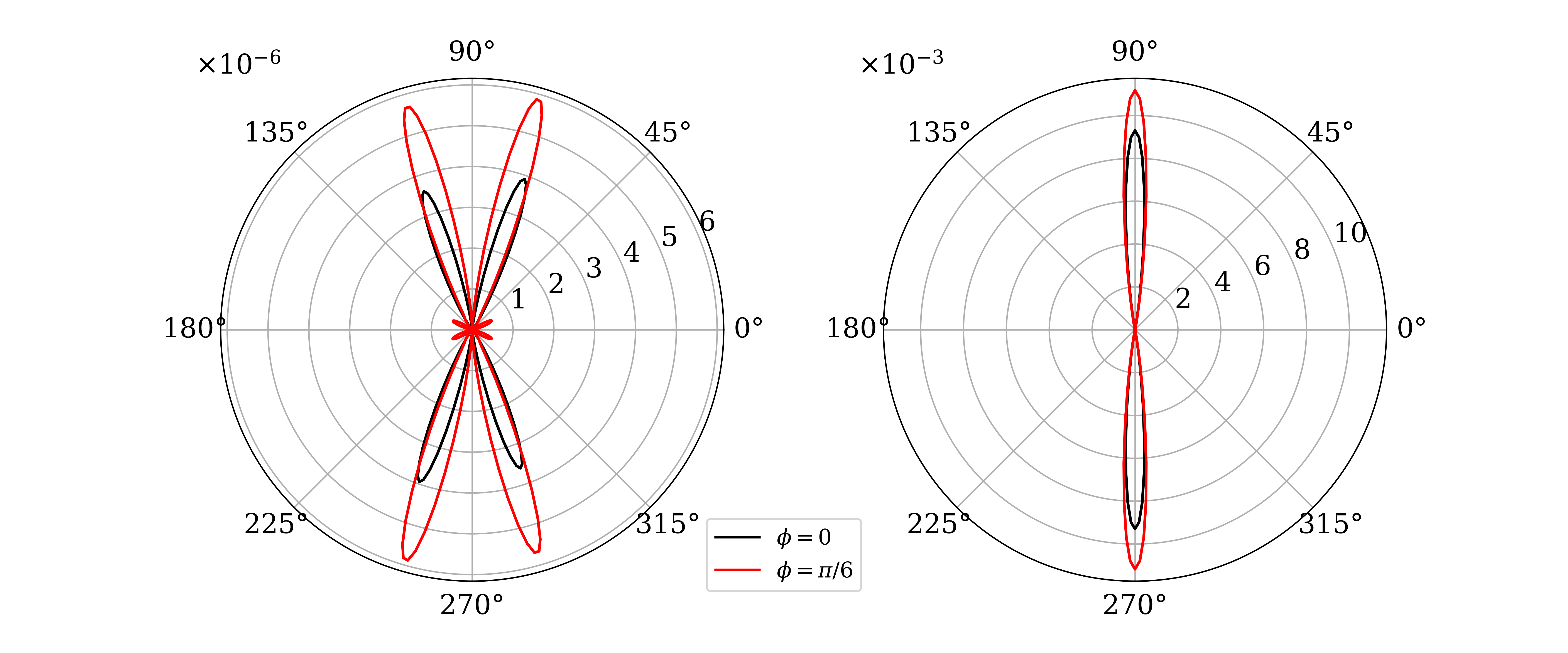}
\caption{Gravitational coupling coefficients as a function of $\beta$ in polar coordinates for different polarization states and azimuthal angles $\phi$. Left: plus polarization, $C_{GW}^{+}$. Right: cross polarization, $C_{GW}^{\times}$.}
\label{fig:TT gauge coupling}
\end{figure}

A dipole antenna is placed inside the cavity to measure the electric field generated by the HFGW. In the case of a background static magnetic field, this electric field is the sum of a contribution from the coupling of the GW to the background electromagnetic field through the inverse Gertsenshtein effect~\cite{Gertsenshtein1962}, and a contribution from the coupling to the mechanical energy, i.e to the cavity walls~\cite{Ratzinger:2024aa}. As we are probing HFGW frequencies that correspond to the cavity first resonances around 5 GHz, which is orders of magnitude above the mechanical resonant frequencies, the cavity is effectively freely falling~\cite{Ratzinger:2024aa}. This means that in the transverse-traceless (TT) gauge, the contribution from the GW tidal forces vanishes and the signal comes entirely from the graviton-photon conversion. 
For $\alpha = 0.1$, the expected superradiant emission persists for 
$\tau_s \simeq 4\,\mathrm{d}$ (see Sec.\,\ref{sec:GWources}). 
Since each tuning step in our data-taking accumulated 7--8\,h of spectra, a monochromatic superradiant signal would remain present over roughly ten consecutive tuning steps throughout the scan. Such a signal would therefore repeatedly reappear in the vertically combined spectra, ensuring that it is not missed despite the stepwise tuning strategy. In addition, self-interaction induces a frequency drift that scales steeply with the gravitational fine-structure parameter, approximately as $\alpha^{17}$~\cite{Aggarwal:2025noe}. For $\alpha \simeq 0.1$, the resulting drift over the 0.01\,s, corresponds to the frequency resolution of 100\,Hz, acquisition time is below 1\,Hz level, far below the resolution bandwidth, so the signal remains confined within a single frequency bin\,\footnote{ However, because of the $\alpha^{17}$ scaling, even a modest increase in $\alpha$ leads to a rapidly growing drift that can exceed the resolution bandwidth, in which case the signal would no longer appear strictly monochromatic.}. In this case, the conversion power in the TT gauge can be calculated as
\begin{equation}
    P_{\text{sig}} = \frac{g}{1+g}\frac{\omega_{G}}{2\mu_{0}} h_{\times,+}^{2} \langle B_{0}^{2} \rangle Q_{l} C_{GW}^{\times,+}(\beta) V\,,
    \label{eq: TT power}
\end{equation}
where $\omega_{G}$ is the angular frequency of HFGW which is matched to the resonant angular frequency of the cavity ($\omega_{c}=2\pi\nu_{c}$), $h_{\times,+}$ are the strains of the HFGW in two different polarizations, $Q_{l}$ is the loaded Q-factor of the cavity, $\sqrt{\langle B_{0}^{2}\rangle}=9.8\,\rm{T}$ is the root-mean-square of the static magnetic field averaged over the cavity volume $V$, and $C_{GW}^{\times,+}(\beta)$ is the form factor for each GW polarization. The form factor is defined as
\begin{equation}
C_{GW}^{\times,+}(\beta,\phi) = \frac{\left|\int d^{3}x\, \vec{E}_{n}^{*}(x) \cdot \hat{j}^{TT}_{\text{eff}} (x,\beta,\phi) \right|^2}{\int |\vec{B}_{0}(x)|^{2}dV\int dV\, \epsilon_{r}(x) |\vec{E}_{n}(x)|^{2}}\,,
\label{eq:GWcoupling}
\end{equation}
where $\vec{E}_{n}$ denotes the electric field profile of the mode $n$ and $\hat{j}_{\rm{eff}}^{TT}$ is the normalized effective current $\vec{j}_{\mathrm{eff}}^{TT}$ induced by GW in the presence of the static magnetic field $\vec{B}_{0}$. The normalized effective current is defined as $\hat{j}_{\rm{eff}}^{TT} \equiv \vec{j}^{TT}_{\rm{eff}} / \omega_{G}h_{\times,+}$.
 Detailed derivation of the conversion power and form factor is provided in Appendix~\ref{app:A}.
GW induce an effective current not only along the axis of the magnetic field, but also along the transverse directions due to the tensor property of GW. The full expressions are given in the Appendix.
The parameter $\beta$ is the angle between the GW propagation direction, which is placed in the $y'-z$ plane, and the cavity axis ($z$ axis) as shown in Fig.\,\ref{fig:GW_cav_angle}.
The $y'-z$ plane is the  $y-z$ plane rotated by $\phi$ about $\hat{z}$, which must be defined since the MC cavity has a partition that breaks the azimuthal symmetry. Therefore, the form factor depends on both $\beta$ and $\phi$. In the calculation of the form factor, one of the three partitions of the MC cavity is aligned along the $\hat{y}$ direction. For a given electric field mode distribution, the $C_{GW}^{\times,+}(\beta,\phi)$ can be evaluated numerically, and  Fig.\,\ref{fig:TT gauge coupling} shows the result at different angles between the partition and the $y-z$ plane.
The CAPP–12T uses the lowest TM mode, whose dominant electric field is oriented along $\hat{z}$ and is uniform along the $z$-axis. As a result, the dominant GW coupling is the cross-polarization, $C_{GW}^{\times}(\beta,\phi)$, which is is four orders of magnitude larger than plus-polarization $C_{GW}^{+}(\beta,\phi)$. 

\begin{figure}[h!]
\centering
    \includegraphics[width=0.7\linewidth]{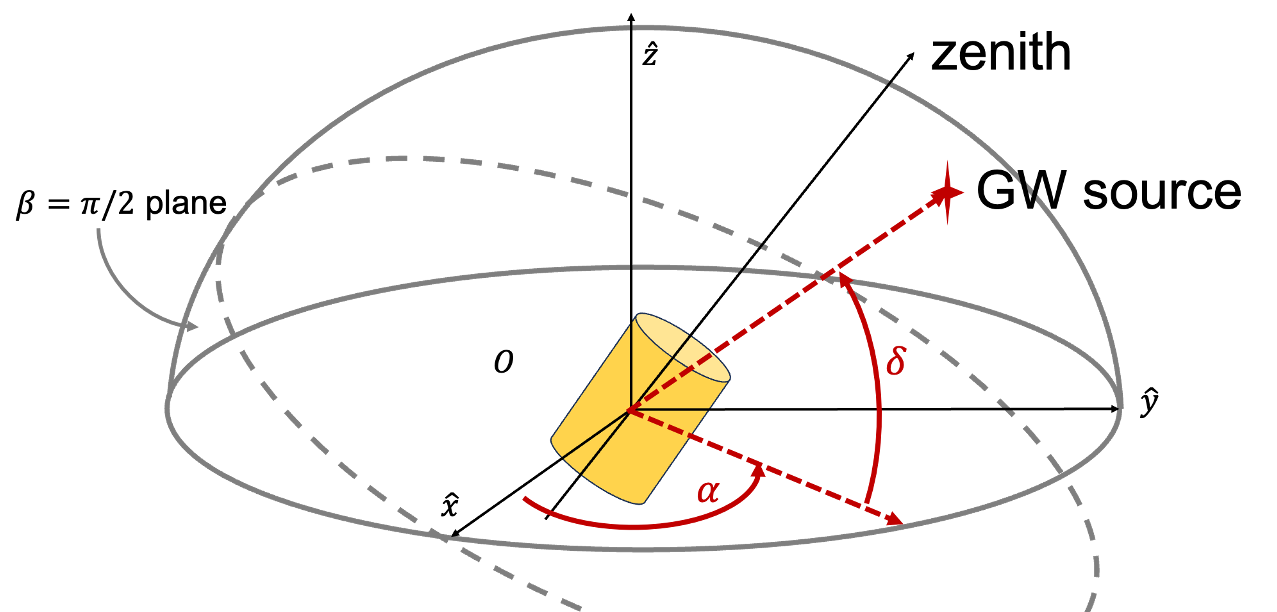}
    \caption{Relative location of the GW source and the cavity in the equatorial coordinates. The GW source position is specified by right ascension $\alpha$ and declination $\delta$, with the zenith direction along $\hat{z}$. The dashed curve represents the $\beta=\pi/2$ plane. }
    \label{fig:GW_cav_angle2}
\end{figure}

\begin{figure*}[htbp]
\centering
    \includegraphics[width=\linewidth]{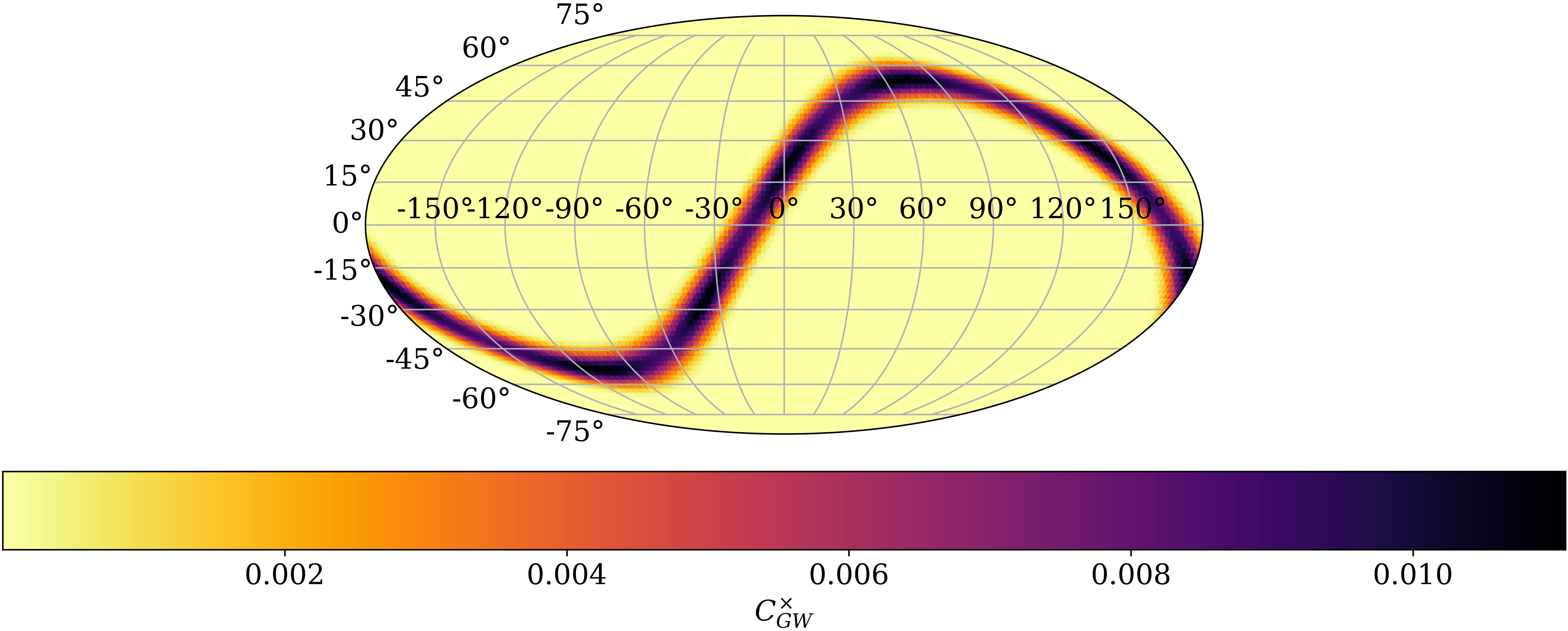}
    \caption{Gravitational-wave with cross polarization coupling coefficients in equatorial coordinates for the CAPP–12T MC cavity mode, evaluated for arbitrary gravitational-wave sources.}
    \label{fig:GW_coupling_sky}
\end{figure*}

Unlike in cavity axion experiments, GW coupling depends on the incident angle of the gravitational wave. This means that the detector sensitivity varies depending on the direction of the gravitational wave source on the celestial sphere. Therefore, one must calculate how the coupling of the resonator changes over time for a source located at an arbitrary right ascension $\alpha$ and declination $\delta$ in equatorial coordinates. As illustrated in Fig.\,\ref{fig:GW_cav_angle2}, the direction of a GW source can be expressed in the equatorial coordinates using the two coordinates $\alpha$ (right ascension) and $\delta$ (declination). The cavity (yellow cylinder) is oriented with its symmetry-axis pointing toward the zenith, along which the external magnetic field $B_{0}$ is applied. The $\beta = \pi/2$ plane is tilted relative to the equatorial plane ($x$--$y$) by the latitude of the detector; in the case of Daejeon, where the CAPP-12T MC experiment was conducted, this tilt is about $36.35^{\circ}$. In short, the latitude of the detector sets the tilt of the cavity axis, and the time dependence of the coupling must be considered as the GW source direction ($\alpha, \delta$) changes relative to this orientation.  
In this coordinate system, GW coupling was numerically calculated for an arbitrary source on the celestial sphere, labeled as GW source in Fig.\,\ref{fig:GW_cav_angle2}, for the two polarization states. Figure\,\ref{fig:GW_coupling_sky} shows the GW coupling for the cross polarization projected onto the celestial sphere, and the result for the plus polarization is shown in the Appendix. The calculation was performed with reference to 09:00:00 KST, and as time progresses, the coupling band shifts along right ascension $\alpha$.
Accordingly, the maximum power induced by the gravitational wave with strain $h_\times = 10^{-20}$ is:
\begin{equation}
\begin{aligned}
    P_{sig} =& 1.83\times 10^{-23} [\mathrm{W}]\left(\frac{h_{\times}}{10^{-20}}\right)^2\left(\frac{C_{GW}}{ 0.012}\right)\\
    &\times \left(\frac{\nu_{G}}{5.311\,\mathrm{GHz}}\right)\left(\frac{\sqrt{\langle B_{0}^{2}\rangle}}{9.8\,\mathrm{T}}\right)^{2}\\
    &\times\left(\frac{Q_{l}}{13,000}\right)\left( \frac{V}{1.4\,\mathrm{L}}\right)\,,
\end{aligned}
\end{equation}
for the CAPP-12T MC experiment.
\section{Data analysis}
\begin{figure*}[htbp]
\centering
    \includegraphics[width=.9\textwidth]{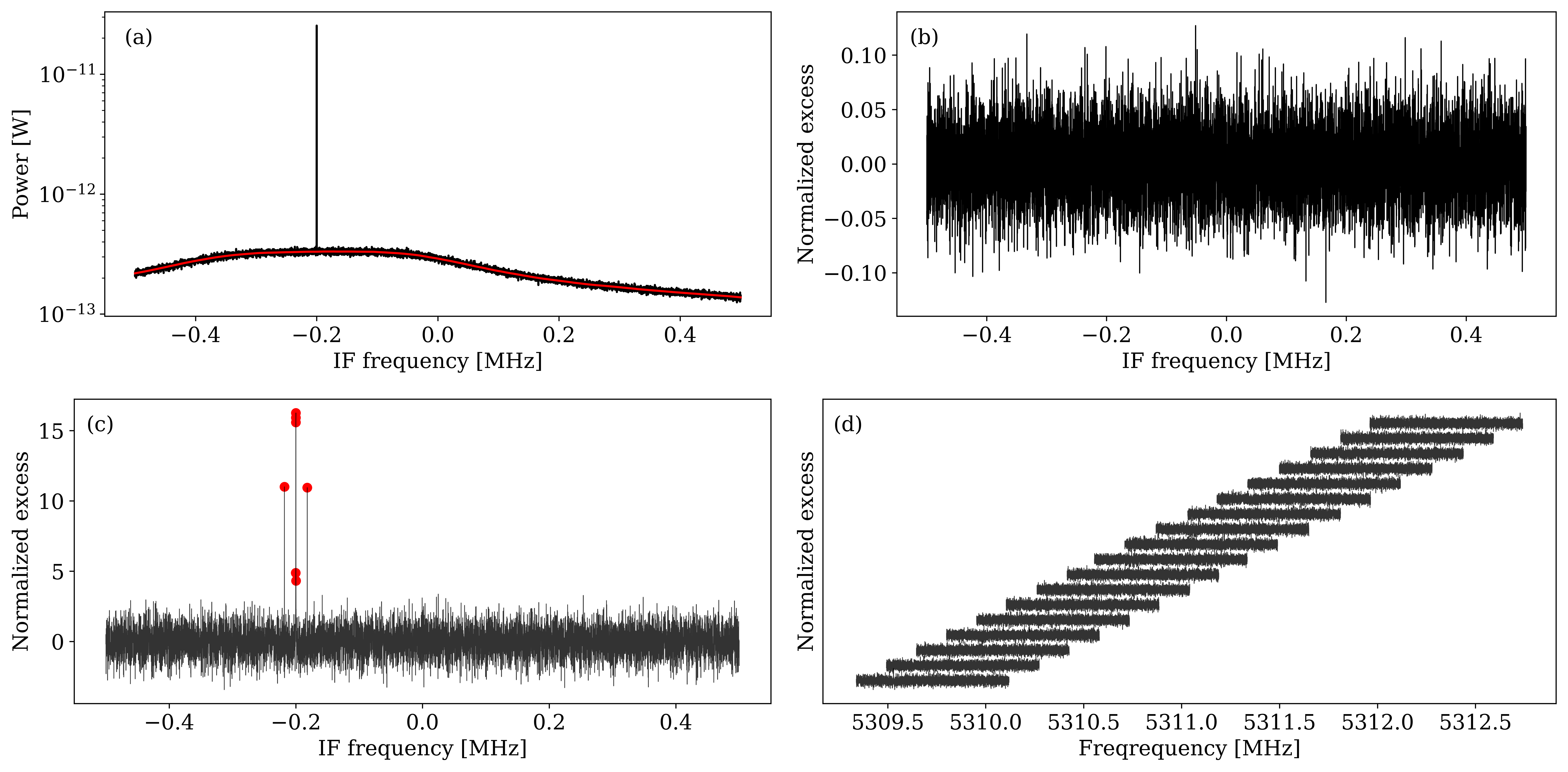}
    \caption{Analysis procedure. (a) Raw spectrum with its baseline (red solid line). The peak at –200\,kHz corresponds to the JPA degenerate mode, which was removed from the analysis. (b) Normalized excess follows a Gaussian distribution centered at zero with a standard deviation of $1/\sqrt{N_{\text{avg}}}\approx0.03$, corresponding $N_{\rm{avg}} = 1000$. (c) After the rescaling process, each 8\,min spectrum obtained during the system characterization procedure was averaged, and the JPA interference peaks [marked in red in (c)] appearing symmetrically around the JPA resonance (–200\,kHz) were removed. (d) The preprocessed spectra and its rescaled spectra with the same cavity resonance frequency were combined using inverse-variance weighted summation. The spectra shown in the figure correspond to data obtained at different resonance frequencies. Here the source of GW is assumed at $(0,0)$ in equatorial coordinates.}
    \label{fig:anal_porcess}
\end{figure*}

The analysis procedure for monochromatic GW follows the methodology outlined in Refs.\,\cite{Ben2018_PRD,CAPP2024_12TMC}; details of the analysis are described in the Appendix. 
First, the raw data are preprocessed to remove spurious peaks that may arise from the JPA degenerate mode (a monochromatic peak in Fig.\,\ref{fig:anal_porcess} (a)) and JPA-induced interference near the JPA resonance (marked in red in Fig.\,\ref{fig:anal_porcess} (c)). A Savitzky-Golay (SG) filter is then applied to fit and subtract the baseline, after which the spectra are divided by the baseline to obtain the normalized power excess, $\delta_n$ [Fig.\,\ref{fig:anal_porcess} (b)]. This $\delta_{n}$ and its deviation are rescaled according to the expected signal-to-noise ratio (SNR) to obtain the spectrum in units of SNR. Subsequently, the 10\,s individual spectra are combined using an inverse-variance weighted summation such that, even for spectra taken at different cavity resonance frequencies, those sharing identical frequency bins are combined (Fig.\,\ref{fig:anal_porcess} (c, d)). This procedure, referred to as \textit{vertical combination}~\cite{Ben2018_PRD}, enhances the statistical significance .

\begin{figure*}[htbp]
\centering
    \includegraphics[width=1\linewidth]{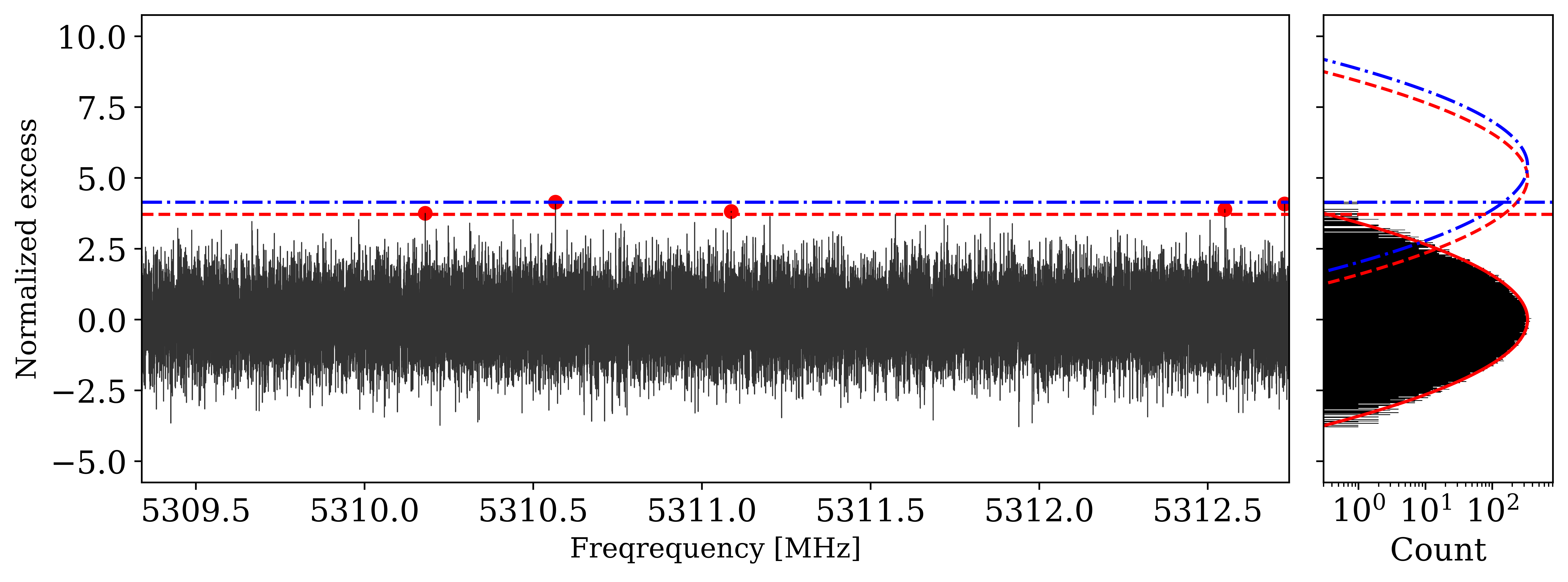}
    \caption{(Left) Vertical grand spectrum for a gravitational-wave source located at $(0,0)$ in equatorial coordinates. 
    The red markers indicate frequency bins above the exclusion threshold, and the blue dashed line shows the sensitivity adjusted to the maximum observed excess. 
    (Right) Histogram of the grand spectrum with the expected distributions overlaid.}
    \label{fig:GrandSpec}
\end{figure*}

The resultant ``grand'' spectrum is then subjected to hypothesis testing to search for possible gravitational wave signatures, as shown in Fig.\,\ref{fig:GrandSpec}, which displays the spectrum assuming a GW source at (0,0) in equatorial coordinates with strain $h_{\times} = 10^{-20}$ (left) and its pull distribution (right).
The pull distribution of the grand spectrum follows a standard normal distribution $\mathcal{N}$(0, 1), shown as the red solid line in the right panel. 
Five outliers, marked as red dots in the figure, are above the threshold value of 3.718 (red dashed line), which corresponds to the 90\% confidence level for the null hypothesis of a $5\sigma$ signal; the corresponding expected distribution is shown as the red dashed curve in the right panel. 
Since rescans were not possible with the current data of these outlier frequency bins, the SNR threshold in this sky direction was adjusted upward to render these outliers to be within the threshold.
For this particular example, the resulting threshold is 5.36.
This adjustment is indicated by the blue dotted-dashed line in the left panel, with the associated expected distribution shown as the blue dashed curve in the right panel.
The same procedure is applied to all sky directions in the full analysis discussed in the next section.
\section{Sensitivity}
\begin{figure*}[htbp]
\centering
    \includegraphics[width=0.9\textwidth]{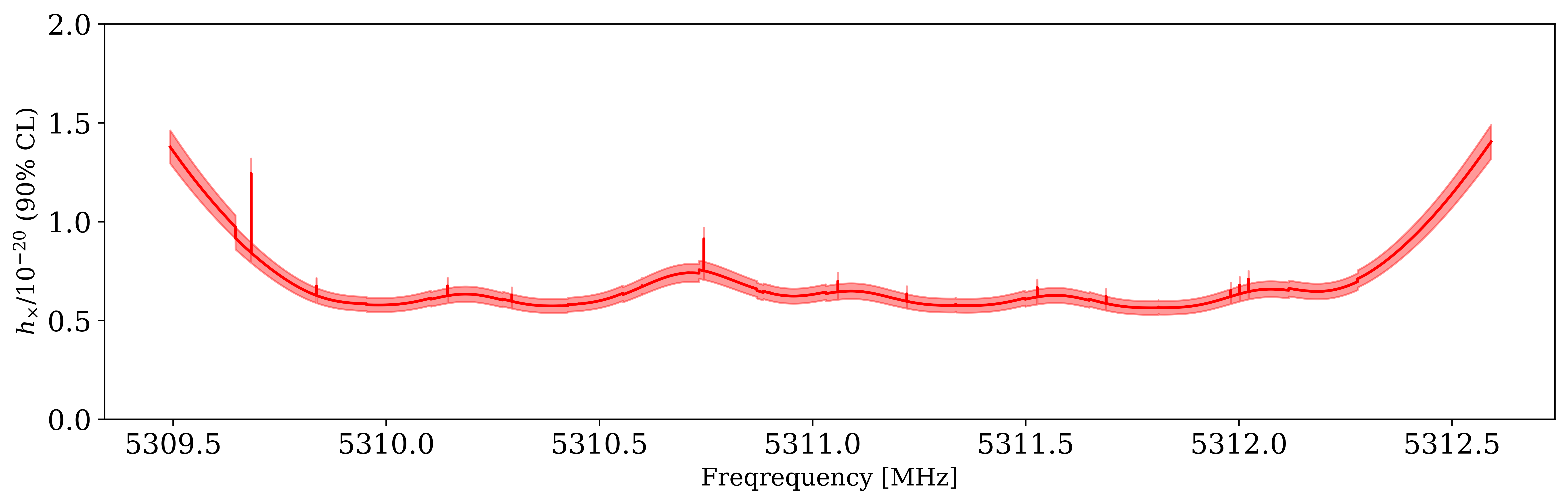}
    \caption{Exclusion limit on the GW strain amplitude $h_{\times}$ at 90\% confidence level around $5.311\,\mathrm{GHz}$. The red curve shows the sensitivity as a function of frequency, and the shaded band indicates the $1\sigma$ total uncertainty obtained by combining all systematic and statistical contributions in quadrature. The small peaks and discontinuities in the exclusion curve arise from experimental conditions and instrumental effects.}
    \label{fig:sensitivity example}
\end{figure*}

The parameter that can be excluded at frequencies where the null hypothesis is rejected is expressed as
\begin{equation}
h_{\times}^{\mathrm{exclude}} = \sqrt{\eta_{\mathrm{SG}} \frac{\mathrm{SNR}^{\mathrm{target}}}{\mathrm{SNR}_{\mathrm{m}}}} h_{\times}^{\mathrm{target}}\,,
\end{equation}
where $\mathrm{SNR}^{\mathrm{target}}$ is the target SNR corresponding to the null-hypothesis signal, which is 5.36 for a GW source located at $(0,0)$ in equatorial coordinates; $\mathrm{SNR}_{\mathrm{m}}$ is the measured SNR at each frequency bin; $h_{\times}^{\mathrm{target}} = 10^{-20}$ is the target GW strain; and $\eta_{\mathrm{SG}}$ denotes the efficiency of the SG filter.
Since baseline fitting with the SG filter may attenuate a gravitational-wave signal, we quantified this effect by software-injecting test signals. 
Using Monte Carlo simulations based on the experimental baseline and the SG filter parameters (window size of 1101 and polynomial order of 4, which is used in~\cite{CAPP2024_12TMC}), we obtained an efficiency of $\eta_{\mathrm{SG}} = 0.999 \pm 0.001$, indicating that any attenuation of the injected signal is negligible. A detailed estimation of the SG filter efficiency is provided in the Appendix.
To determine the overall sensitivity of the experiment, we assumed the GW source at (0,0) in equatorial coordinates is then evaluated by incorporating the updated target SNR and the SG filter efficiency, as shown in Fig.\,\ref{fig:sensitivity example}.

Various sources of uncertainty were considered in determining the exclusion limits. 
The largest contribution arises from the measurement of the noise temperature, amounting to 8.8\% of the combined systematic and statistical errors. 
The second contribution is associated with the GW coupling, originating from the unknown orientation of the cavity wall partition relative to the $y'$–$z$ plane, and is estimated to be 7.6\%. 
Since the coupling coefficient depends on the azimuthal angle $\phi$, this misalignment leads to an uncertainty quantified by the standard deviation of the dependence $\phi$. 
The third contribution originates from the Smith circle fitting~\cite{Pozar:882338} used to determine the antenna coupling, which introduces a fitting error of 1.6\%. 
The final contribution is the statistical fluctuation in the measurement of the loaded factor $Q_{l}$, with typical values around 85 out of 13,000 in this frequency region, which corresponds to an uncertainty of 0.7\%. 
Taking the quadratic sum of all contributions, the total uncertainty amounts to 6.0\% in the strain \footnote{The uncertainties listed in the text represent the ratio of each component defined in terms of the power SNR. Therefore, when calculating the uncertainty in strain, each contribution is weighted by one half.}. 
The red band in Fig.\,\ref{fig:sensitivity example} represents the $1\sigma$ uncertainty band corresponding to this total error.  

\begin{figure*}[htbp]
\centering
    \includegraphics[width=1.\textwidth]{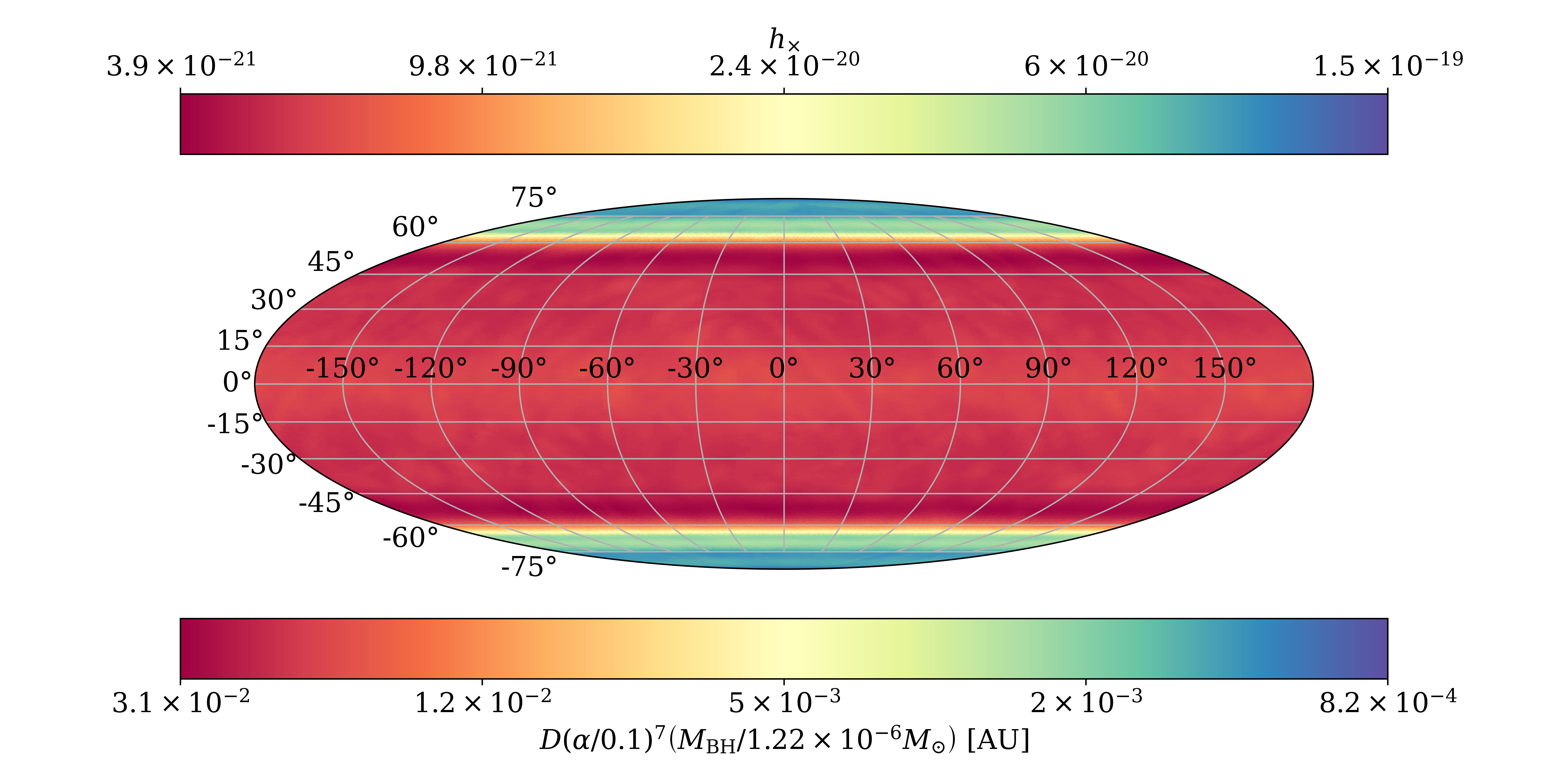}
    \caption{Averaged $h_{\times}$ exclusion (90\% C.L.) over a 2\,MHz span for monochromatic gravitational waves across the sky, and corresponding exclusion distance  for a monogromatic GW at 5.311\,GHz, interpreted in the black-hole superradiance scenario.
    The color scale in the bottom indicates the maximum distance at which a black hole of mass $M_{\rm BH}=1.22\times10^{-6}M_{\odot}$ would be excluded, scaled by $\alpha = 0.1$. In the most sensitive mid-latitude regions, black holes of this mass are excluded within $\mathcal{O}(10^{-2})$\,AU from Earth.}
    \label{fig:CAPP 12T sensitivity}
\end{figure*}

The result shown in Fig.\,\ref{fig:sensitivity example} corresponds to the analysis of a monochromatic GW source assumed to be at $(0,0)$ in equatorial coordinates. 
The expected conversion power depends on both the sky position of the source and the observation time, so the effective weighting of each frequency bin is sky-dependent. Therefore, the analysis must be repeated across a grid of sky positions covering the celestial sphere.
By averaging the sensitivity around $5.311~\mathrm{GHz}$ within a $2~\mathrm{MHz}$ bandwidth over these positions, the results can be projected onto the sky map, as shown in Fig.\,\ref{fig:CAPP 12T sensitivity}.
While the direction of maximum sensitivity varies with time due to Earth’s rotation, as shown in Fig.\,\ref{fig:GW_coupling_sky}, averaging over one sidereal day results in an effective sensitivity band centered around the celestial equator. 
The detector thus exhibits nearly uniform sensitivity for sources within $|\delta| < 60^\circ$, whereas the coupling to gravitational waves from polar regions remains intrinsically weaker and is largely unaffected by rotation. This sky-dependent sensitivity map can be directly interpreted in specific physical scenarios, such as black hole superradiance (Sec.~\ref{sec:GWources}). The reanalyzed frequency range of 2\,MHz centered at 5.331\,GHz corresponds to a black hole mass in units of solar mass $M_{\odot}$ of $1.22\times10^{-6}M_{\odot}$ for $\alpha =0.1$. Then, the corresponding exclusion distance from Earth can be expressed as 
\begin{equation}
    D \approx 1.2\times 10^{-2}\,\mathrm{AU} \left(\frac{10^{-20}}{h_{\times}}\right)\left(\frac{\alpha}{0.1}\right)^{7}\left(\frac{M_{\rm{BH}}}{1.22\times 10^{-6}M_{\odot}}\right).
\end{equation}
The bottom bar of Fig.\,\ref{fig:CAPP 12T sensitivity} represents the corresponding 90\% confidence-level exclusion distance $D$ for a black hole of mass $M_{\rm BH}=1.22\times10^{-6}M_{\odot}$ under the superradiance scenario.
In each sky direction, a black hole of this mass located within $D$ of the Earth would have produced a detectable monochromatic gravitational-wave signal; therefore, our result excludes the existence of such black holes within that distance range. 
For the most sensitive mid-latitude regions, we assume the existence of an axion with the corresponding mass on the order of $11\,\mu\rm{eV}$ matched to the size of the black hole with a mass on the order of $10^{-6}M_{\odot}$ radiating gravitational waves around 5\,GHz (see Eq.\,\ref{eq:ma_superradiance}). Assuming that the primordial black holes were in the superradiant phase during the measurement period, black holes of mass $M_{\rm BH}=1.22\times10^{-6}M_{\odot}$ are excluded within a distance of $D = (3.2\pm0.2)\times 10^{-2}\,\mathrm{AU}$ from Earth.

\section{Conclusion}

We reanalyzed data from the CAPP–12T MC haloscope experiment to search for monochromatic high-frequency gravitational waves in a 2\,MHz band around 5.311\,GHz. 
We set 90\% confidence-level exclusion limits on the strain amplitude, reaching $h_0 \sim 3.9\times10^{-21}$ in the most sensitive regions. 
Within the black hole superradiance scenario, assuming the existence of an axion with a mass around $2.7\,\mu\rm{eV}$, these limits exclude black holes of mass $M_{\rm BH}\simeq 1.22\times10^{-6}M_{\odot}$ within a distance of $\mathcal{O}(10^{-2})$\,AU from the Earth
While the present interpretation focuses on the two-axion annihilation channel, the accessible parameter space can vary substantially depending on the underlying gravitational-wave production mechanism, and alternative models may yield different signal morphologies or frequency ranges.
This work demonstrates that resonant cavities with quantum amplifiers can serve as sensitive detectors of well-motivated HFGW sources.
Although the present form factors are not optimized, future dedicated setups could achieve substantially higher (estimated on the order of a factor of three) coupling and thus improve the attainable sensitivity.
Future searches may extend to broader frequency ranges and probe transient signals, such as those from primordial black holes.

\begin{acknowledgments}
The authors acknowledge helpful discussions with Pedro Schwaller and Joachim Kopp.
This work has been supported by the Cluster of Excellence “Precision Physics, Fundamental Interactions, and Structure of Matter” (PRISMA++ EXC 2118/2) funded by the German Research Foundation (DFG) within the German Excellence Strategy (Project ID 390831469),
and by the European Union (GravNet, ERC-2024-SyG 101167211, DOI: 10.3030/101167211). Views and opinions expressed are however those of the author(s) only and do not necessarily reflect those of the European Union or the European Research Council Executive Agency. Neither the European Union nor the granting authority can be held responsible for them.
This work was also supported by the Institute for Basic Science (IBS-R017-D1-2024-a00 and IBS-R040-C1-2025-a00) and by JSPS KAKENHI (Grant No.~JP22H04937).
The work of Y. Kim was supported by the Alexander von Humboldt Foundation. 
J. Jeong was partially supported by the Knut and Alice Wallenberg Foundation.
A. F. Loo was supported by a JSPS Postdoctoral Fellowship.
J. E. Kim was partially supported by the Korea Academy of Science.
This publication is part of the R\&D\&i project PID2023-146686NB-C31 funded by MICIU/AEI/10.13039/501100011033/ and by ERDF/EU.
IFAE is partially funded by the CERCA program of the Generalitat de Catalunya.
Diego Blas acknowledges the support from the European Research Area (ERA) via the UNDARK project of the Widening participation and spreading excellence programme (project number 101159929).
Jordan Gué is funded by Grant No. CNS2023-143767, funded by MICIU/AEI/10.13039/501100011033 and by European Union NextGenerationEU/PRTR
\end{acknowledgments}
\appendix
\section{Conversion power}\label{app:A}

In this appendix, we derive the resonant conversion power of a gravitational wave (GW) in a cavity via the inverse Gertsenshtein effect.
A static magnetic field $B_{0}$ is applied in the $z$-direction and is assumed to be spatially uniform throughout the cavity. The effective current arising from the GW–magnetic-field interaction is~\cite{Berlin2021},
\begin{equation}
    j^{\mu}_{\rm eff} = \partial_{\nu}\!\left(\tfrac{1}{2}hF^{\mu\nu} + h^{\nu}_{\alpha}F^{\alpha\mu}-h^{\mu}_{\alpha}F^{\alpha\nu}\right),
    \label{eq:effecitve_current_full}
\end{equation}
where $h_{\mu\nu}$ is the GW strain tensor (and its index-raised forms appearing in Eq.\,\eqref{eq:effecitve_current_full}) and $F^{\mu\nu}$ is the electromagnetic field-strength tensor.
For simplicity, we consider a GW propagating in the direction $\hat{n}=\hat{z}'$, which lies in the $y$–$z$ plane. In the GW-propagation frame (TT gauge), the strain tensor is~\cite{Schutz:1985jx},
\begin{equation}
    h =
    \begin{pmatrix}
    h_{+} & h_{\times} & 0 \\
    h_{\times} & -h_{+} & 0 \\
    0 & 0 & 0
    \end{pmatrix}
    e^{-i\omega_{G}(t-z')},
\end{equation}
where $h_{+,\times}$ are the plus and cross polarizations, respectively, and $\omega_{G}$ is the GW angular frequency.

The strain tensor for a wave propagating in the direction $\hat{n} = \cos\beta\, \hat{z} + \sin\beta\, \hat{y}$ can be obtained by a rotation matrix about the $x$-axis:
\begin{equation}
R(\beta) =
    \begin{pmatrix}
    1 & 0 & 0 \\
    0 & \cos\beta & \sin\beta \\
    0 & -\sin\beta & \cos\beta
    \end{pmatrix},
\end{equation}
and the strain tensor in the lab basis is
\begin{equation}
\begin{aligned}
    h^{\rm TT}_{\hat{n}} =&
    \begin{pmatrix}
    h_{+} & h_{\times}\cos\beta & -h_{\times}\sin\beta \\
    h_{\times}\cos\beta & -h_{+}\cos^{2}\beta & h_{+}\cos\beta\sin\beta \\
    -h_{\times}\sin\beta & h_{+}\cos\beta\sin\beta & -h_{+}\sin^{2}\beta
    \end{pmatrix}\\
    &\times
    e^{-i\omega_{G}\!\left(t - (\sin\beta\, y + \cos\beta\, z)\right)}.
\end{aligned}
\end{equation}

For a multi-cell cavity, we use a mode with the electric field only along $z$, $E_{z}$; hence, the effective current has only a $z$-component. The $z$-component of the effective current is
\begin{equation}
    j^{3}_{\rm TT} = \left( \partial_{1} h^{3}{}_{2} - \partial_{2} h^{3}{}_{1} \right) B_{0}.
\end{equation}
Since the strain tensor $h$ has no $x$-dependence, only the second term contributes:
\begin{equation}
    j^{3}_{\rm TT} = -i\, h_{\times}\, \omega_{G}\, B_{0}\, \sin^{2}\beta\, e^{-i\omega_{G}\!\left(t - (\sin\beta\, y + \cos\beta\, z)\right)}.
\label{eq:TT_current_3}
\end{equation}

The wave equation for the electric field, including the leading-order effective current induced by the gravitational wave, is
\begin{equation}
    \nabla \times \nabla \times \vec{E} - \partial_{t}^{2} \vec{E} = -\partial_{t}\, \vec{j}_{\text{eff}}.
\end{equation}
In the TT gauge, at high frequencies where the cavity is approximately freely falling, the general solution for the electric field can be written as a modal expansion~\cite{Jackson:1998nia}:
\begin{equation}
    \vec{E}(t, \vec{x}) = \sum_{n} e_{n}(t)\, \vec{E}_{n}(\vec{x}).
\end{equation}
In the steady state, the measured signal is
\begin{equation}
    \vec{E}_{\text{sig}}(t, \vec{x}) =
    - \frac{\displaystyle \int dV\, \vec{E}_{n}^{*} \cdot \vec{j}^{\rm TT}_{\text{eff}}}
           {\displaystyle \int dV\, \epsilon_{r}\, |\vec{E}_{n}|^{2}}\,
      \frac{Q_{m}}{\omega_{G}}\, \vec{E}_{n}(\vec{x}),
\label{eq:TT_Esol}
\end{equation}
where $Q_{m}$ is the reduced quality factor given by the harmonic sum of the cavity quality factor $Q_{c}$ and the gravitational-wave coherence quality factor $Q_{g}$. For monochromatic gravitational waves, we approximate $Q_{m} \approx Q_{c}$. Here, $\epsilon_{r}$ is the relative permittivity inside the cavity, introduced by the presence of a dielectric tuning rod.

The GW-to-EM conversion power in the cavity is~\cite{Jackson:1998nia},
\begin{equation}
    P_{\text{con}} = \omega_{c}\, \frac{U}{Q_{c}}
    = \frac{\omega_{c}}{2}\, \frac{\displaystyle \int dV\, \epsilon_{r}\, |\vec{E}_{\text{sig}}|^{2}}{Q_{c}}
    \Big|_{\omega_{c} = \omega_{G}}.
\label{eq:conv_p}
\end{equation}

Substituting Eqs.\,\eqref{eq:TT_current_3} and \eqref{eq:TT_Esol} into Eq.\,\eqref{eq:conv_p}, we obtain
\begin{widetext}
\begin{equation}
    P_{\text{con}} = \frac{\omega_{G}}{2}\, h_{\times}^{2}\, B_{0}^{2}\, Q_{c}\, V\,
    \left(
      \sin^{4}\beta\,
      \frac{\left| \displaystyle \int dV\, \vec{E}_{n}^{*} \cdot \hat{z}\, e^{i\omega_{G}(\sin\beta\, y + \cos\beta\, z)} \right|^{2}}
           {\displaystyle V \int dV\, \epsilon_{r}\, |\vec{E}_{n}|^{2}}
    \right).
\end{equation}
\end{widetext}
The bracketed term is commonly referred to as the form factor in axion haloscope experiments (often denoted $C$ or $G$); here, we refer to it as the GW form factor, $C_{GW}(\beta)$. Thus, the conversion power simplifies to
\begin{equation}
    P_{\text{con}} = \frac{\omega_{G}}{2}\, h_{\times}^{2}\, B_{0}^{2}\, Q_{c}\, C_{GW}(\beta)\, V,
\end{equation}
which is Eq.~\ref{eq: TT power}.

For completeness, the other components of the effective current are
\begin{align}
    j^{1} &= \left(\tfrac{1}{2}\partial_{2}\mathrm{Tr}(h) + \partial_{0}h_{02}
              -\partial_{2}(h_{22}+h_{11}) - \partial_{3}h_{32} \right)B_{0},\\
    j^{2} &= -\left(\tfrac{1}{2}\partial_{1}\mathrm{Tr}(h) + \partial_{0}h_{01}
              - \partial_{3}h_{31} - \partial_{1}(h_{11}+h_{22})\right)B_{0},
\end{align}
and, for the geometry in Fig.~\ref{fig:GW_cav_angle}, these reduce to
\begin{align}
    j^{1}_{\rm TT} &= -i\,\omega_{G}\,\sin\beta\, h_{+}\,
    e^{-i\omega_{G}\!\left(t - (\sin\beta\, y + \cos\beta\, z)\right)} B_{0},\\
    j^{2}_{\rm TT} &= -i\,\omega_{G}\,\sin\beta\cos\beta\, h_{\times}\,
    e^{-i\omega_{G}\!\left(t - (\sin\beta\, y + \cos\beta\, z)\right)} B_{0}.
\end{align}
These currents contribute to the gravitational couplings $C_{GW}^{+,\times}$ in Eq.~\ref{eq:GWcoupling}.

\section{Detailed analysis}
The spectra were preprocessed to remove the JPA degenerate mode located at $-200\,\mathrm{kHz}$ from the cavity resonance, and the corresponding frequency bin was replaced by the average of neighboring bins. In addition, suspicious peaks appearing symmetrically near the JPA resonance due to interference (red points in Fig.\,\ref{fig:anal_porcess} (c); see details below) were also replaced with the average of adjacent bins. These bins were assigned infinite variance in the weighted average calculation, and consequently, in the inverse-variance weighted summation process, they effectively carry zero weight and are excluded from the sensitivity estimation.
For every spectrum collected from 10-second digitizer data, a SG filter with window size 1101 and polynomial degree 4 is applied for baseline fitting as shown in Fig.\,\ref{fig:anal_porcess} (a). Each spectrum is then normalized to obtain the dimensionless spectrum $\delta_{n}$, defined as
\begin{equation}
    \delta_{n} = \frac{\mathrm{spectrum}}{\mathrm{fit}} - 1.
    \label{eq:npe}
\end{equation}
Figure\,\ref{fig:anal_porcess} (b) shows the normalized excess of the spectrum shown in (a). The standard deviation was 0.03, consistent with the expected value of $1/\sqrt{N_{\rm{avg}}}$ , where $N_{\rm{avg}} = 1000$ is number of averages for each frequency bin, given by the product of the total measurement time (10\,s) and the resolution bandwidth (100\,Hz).
The normalized spectrum is then scaled by multiplying by $(P_{\rm exp}/\delta P_{\rm sys})^{-1}$ where $P_{exp}$ is the expected signal power assuming $h_{\times} = 10^{-20}$, and $\delta P_{\rm sys}$ is the system noise power defined as $\delta P_{\rm{sys}} = k_{B}T_{\rm sys}\Delta\nu$ where $k_{B}$ is the Boltzmann constant, $T_{\mathrm{sys}}$ is the measured system noise temperature, and $\Delta \nu$ is the resolution bandwidth.
The expected signal is calculated under the assumption that GW are monochromatic. For a GW with frequency $\omega_{G}$ interacting with a cavity resonant at $\omega_{c}$, the expected signal power at time $t$ from direction $(\phi, \theta)$ is given by
\begin{widetext}
\begin{equation}
    P_{\mathrm{exp}}(\omega_{G}; t, \phi, \theta) = 
    \frac{\beta}{\beta + 1} \frac{\omega_{G}}{2\mu_{0}} h_{\times}^{2} \langle B_{0}^{2}\rangle Q_{l} 
    C_{GW}(\beta(t, \phi, \theta)) V \left(
    \frac{1}{1 + 4 Q_{l}^{2} \left( \frac{\omega_{G}}{\omega_{c}} - 1 \right)^2 }\right).
\end{equation}
\end{widetext}
The system noise temperature $T_{\mathrm{sys}}$ was measured using the SNR improvement method during the cavity and amplifier check procedure conducted every 6 minutes. The result was
$T_{\mathrm{sys}} = 356\,\mathrm{mK} \pm 31\,\mathrm{mK}_{\mathrm{stat.}} \pm 7.2\,\mathrm{mK}_{\mathrm{sys.}}$,
where the systematic uncertainty originates from the discrepancy between the Y-factor method and the NVR method.

After rescaling, the spectra measured over an 8-minute interval were combined at each frequency bin using an inverse-variance weighted average. Figure\,\ref{fig:anal_porcess} (c) shows the result of pre-processing where only the JPA degenerate mode was removed. Symmetric suspicious peaks are observed around the JPA resonance frequency. These peaks were found to appear at the same position in the intermediate-frequency (IF) domain even when the cavity resonance frequency was varied, indicating that they originate from intrinsic interference within the JPA itself. Therefore, these frequency bins were additionally excluded in the preprocessing stage as described above.
The post-processed spectra with the same tuning step (i.e., spectra with the same cavity resonance frequency) are then combined using an inverse-variance weighted average (see Fig.\,\ref{fig:anal_porcess} (d)).

Then, the spectra are vertically combined using inverse-variance weighted summation to minimize the variance increase during aggregation, yielding the power excess $\delta_{v}$ and standard deviation $\sigma_{v}$ for each frequency bin. After this combination, the ratio $\delta_{v}/\sigma_{v}$, the normalized excess follows a standard normal distribution. 

\subsubsection{SG filter efficiency}
White noise was first added to the measured baseline, and a monochromatic GW signal with an excess of 5 was injected. 
The resulting data were then processed with the SG filter to evaluate the degradation in SNR. 
Because the injected excess is expected to follow a normal distribution, we generated 600 independent white-noise samples following the normal distribution. 
For each sample, the full analysis procedure, including SG filter fitting, was performed to obtain the distribution of excess values at the injection frequency.
\begin{figure}[H]
\centering
    \includegraphics[width=.8\linewidth]{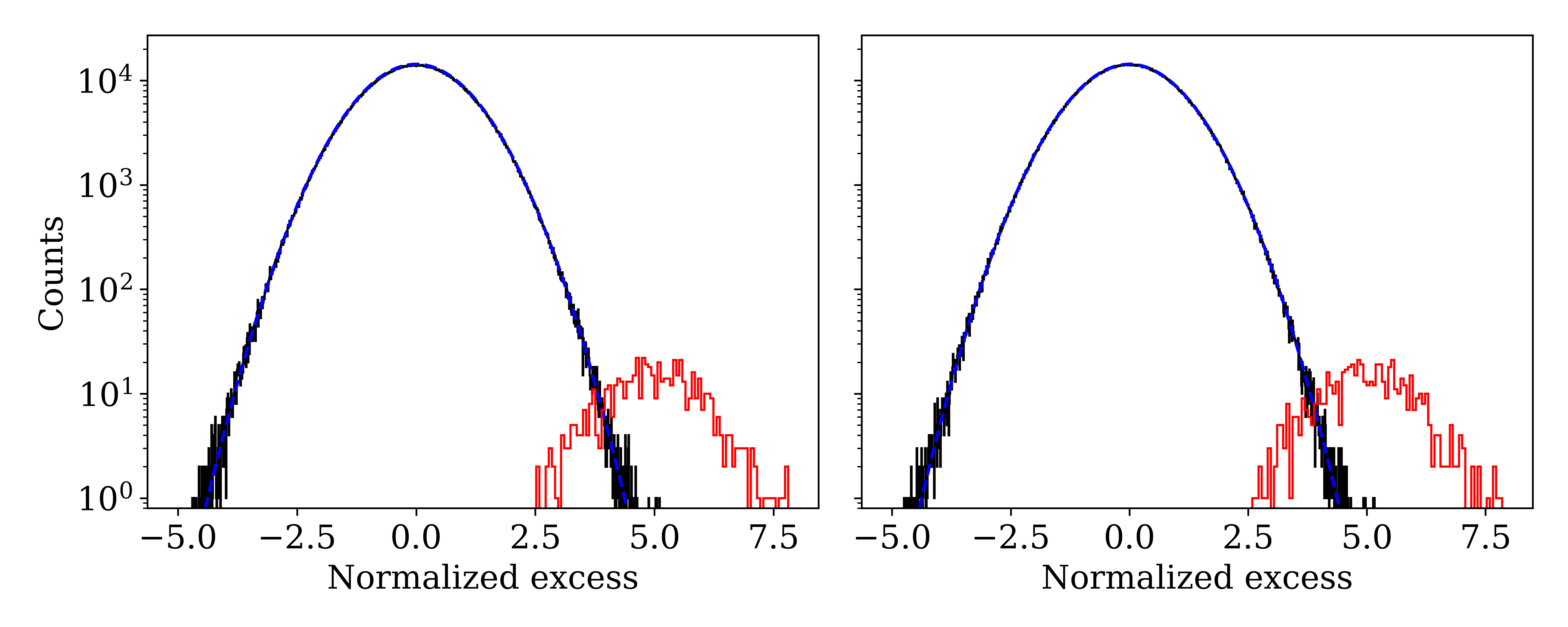}
    \caption{Distributions of normalized excess values obtained from 600 white-noise samples with an injected monochromatic GW signal of excess 5. 
    The black histograms represent the noise-only spectra, and the blue curves are Gaussian fits, both yielding a standard normal distribution $\mathcal{N}(0,1)$. 
    The red histograms correspond to the distribution of normalized excess values from frequency bins where artificial GW signals were injected. 
    (Left) Without baseline fitting (flat baseline), the signal distribution is centered at 5.0. 
    (Right) After applying the SG filter (window size 1101, polynomial order 4), the signal distribution is centered at 4.998.}
    \label{fig:Sg_hists}
\end{figure}
From these distributions, the mean excess was extracted and compared with the injected value of 5, providing a direct measure of SNR degradation for a given baseline. Figure\,\ref{fig:Sg_hists} shows the mean of excess is reduced to 4.998.
This procedure was repeated for all resonance frequencies.
The resulting set of mean degradation values, together with their standard deviations, was used to characterize the overall impact of the SG filter. 
\bibliographystyle{apsrev4-2}
\bibliography{main}

\end{document}